\documentclass[final, times, sort&compress]{elsarticle}
\usepackage{lineno,hyperref}
\usepackage{graphicx}
\usepackage{dcolumn}
\usepackage{bm}
\usepackage{upgreek}
\usepackage{hyperref}
\usepackage{ams math}
\usepackage{multirow}
\usepackage[margin=1.3in]{geometry}

\usepackage{multicol} 

\usepackage{nomencl} 

\let\oldequation\equation
\let\oldendequation\endequation

\renewenvironment{equation}
  {\linenomathNonumbers\oldequation}
  {\oldendequation\endlinenomath}

\usepackage{hyperref}

\usepackage{etoolbox}
\renewcommand\nomgroup[1]{%
  \item[\bfseries
  \ifstrequal{#1}{G}{Greek symbols}{%
  \ifstrequal{#1}{S}{Subscripts}{}}%
]}

\journal{Fuel}

\begin{document}

\begin{frontmatter}

\title{Mist formation during micro-explosion of emulsion droplets}

\author{Houpeng Zhang}
\author{Zhen Lu}
\author{Tianyou Wang}
\author{Zhizhao Che\corref{cor1}}
\cortext[cor1]{Corresponding author.
}
\ead{chezhizhao@tju.edu.cn}
\address{State Key Laboratory of Engines, Tianjin University, Tianjin, 300350, China.}


%
%

\begin{abstract}
The micro-explosion of emulsion droplets plays an important role in promoting atomization, improving combustion efficiency, and reducing pollutant emissions. In this experimental study of the micro-explosion of emulsion droplets, we find that mist can be generated during the heating of emulsion droplets, and the mist generation is closely related to the micro-explosion process. Combined analysis from high-speed images, gas chromatography, and droplet temperature variation shows that the mist generation is due to the condensation of vapor into small droplets as the temperature decreases. Two micro-explosion modes are observed, intense micro-explosion with a large amount of mist and weak micro-explosion with a small amount of mist. Different emulsified fuels are tested, and mist can be produced for all the emulsified fuels. The mist is quantitatively analyzed via digital image processing. According to the mist concentration curve in the micro-explosion process, the micro-explosion mode can be distinguished. The effects of the water and surfactant contents in the emulsion droplets are studied, and the mists are used to characterize the micro-explosion. Increasing the water content can promote the vaporization of the water phase, increase the strength of micro-explosion, and result in a large amount of mist. Increasing the surfactant content can improve the stability of the emulsion droplet, reduce the probability of intense micro-explosion, and hence reduce the mist concentration.
\end{abstract}

\begin{keyword}
\texttt {
Droplet micro-explosion \sep
Mist formation \sep
Emulsified fuel \sep
Water-in-oil emulsion \sep
Multi-component fuel
}
\end{keyword}

\end{frontmatter}


\section{Introduction} \label{sec:sec1}

Efficient combustion is the main problem of liquid fuel utilization in many industries such as transport and power engineering. The solution to these issues could improve the energy efficiency and reduce the negative impact on the environment \cite{Antonov2020TwoComponentDroplets}. Hence, increasing thermal efficiency and reducing pollutant emissions has always been an important task in the development of advanced spray burners \cite{Hoang2021CombustionBehavior, Zhang2018SuperheatLimit}. Multi-component fuels have been adopted by many combustion systems, especially in diesel engines, for various purposes including flame control, higher efficiency, and NOx reduction \cite{MeloEspinosa2017EnginePerformance, Nadeem2006EmissionCharacteristics, Vershinina2016CoalWaterSlurryContaining, Watanabe2010BreakupCharacteristics}. An important advantage of multi-component fuels is that they could induce micro-explosion, which could promote the secondary breakup in fuel atomization processes and hence improve combustion efficiency \cite{Antonov2019ConductiveHeatingRadiantHeating, Shen2020HomogeneousFuel, Yi2017DropletEvaporation}.

Emulsified fuels are multi-component fuels made of water and combustible oil. Because of the difference in the boiling points between water and oil, when emulsified fuels are sprayed into high-temperature combustion chambers, water may boil first and result in the puffing or micro-explosion phenomenon. The puffing and micro-explosion of emulsion droplets can increase the surface area of evaporation and chemical reactions \cite{Lyu2021MutualSolubilityDifferentials, Nyashina2020SlurryFuel}. As a result, it can significantly speed up the fragmentation and vaporization of multi-component fuels and improve combustion efficiency \cite{Moussa2020FluorescenceBiofuels, Mura2012WaterMetastableState, Nyashina2020SlurryFuel, Rosli2021SuspendedDroplet}. To predict the overall atomization performance of multi-component fuels, it is necessary to understand the dynamic behavior and fragmentation characteristics of emulsion droplets. The puffing and micro-explosion phenomenon of emulsion droplets have been extensively studied \cite{Antonov2020ImmiscibleTwoComponentDroplet, Bulat2018ThresholdIntensity, Castanet2022CompositeDroplets, Harada2011SprayCombustion, Khan2014MicroExplosionOccurrence, Sazhin2022MultiComponentFuel, Shen2020EvaporationModel}. If the entire droplet quickly breaks up into a large number of secondary droplets, the phenomenon is called micro-explosion; when water vapor is only blown from the surface of the droplet, accompanied by the ejection of secondary droplets, the phenomenon is called puffing \cite{Avulapati2016DieselBiodieselEthanolBlends}.

At present, there are many experimental studies considering the micro-explosion and puffing of emulsion droplets. The micro-explosion of emulsion droplets highly depends on the emulsion properties, such as the water content, the particle size of the dispersed phase, the viscosity and surface tension of the continuous phase, and surfactant content \cite{Antonov2020DifferentHeatingSchemes, Califano2014SingleDropletStability, Jang2019SurfaceTensionLightAbsorbance, Shen2020HomogeneousFuel, Shen2019AlternativeFuel, Shen2020DropletBreakup, Shinjo2014MicroexplosionPhysics}. Gao et al.\ \cite{Gao2022PolymerSurfactant} showed that the occurrence of breakup modes is affected by the dispersed water diameter, and emulsion droplets with larger dispersed water droplets are more likely to induce micro-explosion. Pavel et al.\ \cite{Strizhak2021ChildDropletsSchlierenPhotography} studied the micro-explosion of two kinds of droplets, i.e., emulsion droplets and two-component immiscible droplets. As the heating temperature increases, the size of the secondary droplets produced by the breakup of the immiscible two-component droplets decreases, and the number of secondary droplets increases; however, the emulsion droplets show the opposite behavior. Wang et al.\ \cite{Wang2022EvaporationCharacteristics} studied the evaporation and micro-explosion of the emulsion droplets with different oils, and found a larger difference in the boiling points of the components and a higher ambient temperature are favorable for puffing and micro-explosion. Madan et al.\ \cite{Avulapati2019SecondaryAtomizationMicroEmulsion} studied the puffing and micro-explosion of diesel-biodiesel-ethanol emulsion droplets. Phase separation of the emulsion constituents was observed within the emulsion droplets before micro-explosion and puffing. A single nucleus of the separated components in the droplet results in a strong micro-explosion, while multiple nuclei result in a weaker micro-explosion. Shen et al.\ \cite{Shen2020DropletBreakup} studied the effect of surfactants on the micro-explosion of the emulsion droplets, and found that after the surfactant at the water/oil interface is deactivated at high temperature, water coalescence occurs, resulting in catastrophic micro-explosion. Huang et al.\ \cite{Huang2020MicroExplosionBlendedDroplet} studied the evaporation and micro-explosion of biodiesel/n-propanol blended droplets. With the increase of the n-propanol concentration, the micro-explosion intensity and evaporation rate of the emulsion droplets increase first and then decrease, while the micro-explosion delay time decreased first and then increases. Melo-Espinosa et al.\ \cite{MeloEspinosa2018MicroChannelEmulsifier} found the time delay of the bubble formation and growth in the emulsion droplets is affected by the surfactant content. For emulsions without surfactants, shorter puffing periods and larger bubbles were observed. Zhang et al.\ \cite{Zhang2018SuperheatLimit} conducted experimental observations of micro-explosion of emulsion droplets under both atmospheric pressure and diesel-like conditions. With the increase of the hydrous ethanol concentration, the intensity of the micro-explosion was found to follow a parabolic relationship, and droplets with 30\% and 40\% hydrous ethanol contents exhibited a higher possibility of complete micro-explosion. Wang et al.\ \cite{Wang2022MechanismAmbientTemperatures} correlated bubble formation with micro-explosion to explore the micro-explosion modes at different ambient temperatures. With the increase in ambient temperature, they found that bubble nucleation in the droplet gradually changed from a single site to multiple sites, and the bubble nucleation sites gradually shifted from the droplet center to the surface.

To further understand the mechanism of the micro-explosion and puffing of emulsion droplets, several simplified models and numerical simulations have been reported. Shen et al.\ \cite{Shen2020EvaporationModel} developed a model to consider the heat/mass transfer for the heating and evaporation of emulsion droplets. Besides the heat transfer and mass transfer inside the droplet and on the droplet surface, the deactivation of the surfactant and the coalescence of water droplets were also considered. Shinjo et al.\ \cite{Shinjo2014MicroexplosionPhysics} investigated the micro-explosion/puffing of water-in-oil emulsion droplets using interface-capturing simulations. They found that the micro-explosion can occur and increase the breakup intensity due to the interaction between multiple puffings. Each explosion initially starts independently as puffing, but soon the entire droplet may break up to a much greater extent due to their mutual interactions. Stavros et al.\ \cite{Fostiropoulos2020HeavyFuelOil} predicted numerically the heating and explosive boiling leading to the fragmentation of heavy fuel oil-water droplets. It was revealed that puffing and micro-explosion can speed up the droplet breakup by almost an order of magnitude relative to the aerodynamic breakup. Watanabe et al.\ \cite{Watanabe2010EmulsifiedFuelSpray} proposed a mathematical model for puffing. The rate of mass change of the droplet during puffing was expressed by the evaporation rate of dispersed water and the mass change rate due to fine droplets spouted from the droplet surface.

The characterization of the micro-explosion intensity is important for analyzing the micro-explosion process. Many studies have been devoted to the micro-explosion intensity  \cite{Avulapati2016DieselBiodieselEthanolBlends, Avulapati2019SecondaryAtomizationMicroEmulsion, Chen2017SuspendedDropletExperimentDropletCombustion, Meng2019BiodieselEthanol, Meng2021DifferentOxygenConcentrations, Moussa2020FluorescenceBiofuels, Wang2022EvaporationCharacteristics}. Meng et al.\ \cite{Meng2019BiodieselEthanol} established a correlation for calculating the micro-explosion intensity of emulsion droplets. According to the intensity, they classified the micro-explosion into four levels, namely slight micro-explosion, medium-intensity micro-explosion, strong micro-explosion, and super micro-explosion. Wang et al.\ \cite{Wang2019NanofuelEvaporation} investigated the evaporation and micro-explosion of cerium-oxide-based nanofuel droplets. They defined a parameter to quantify the micro-explosion intensity, which increases with the nanoparticle concentrations. Wang et al.\ \cite{Wang2019NucleationVaporCloud} examined the effect of ambient temperature on the micro-explosion of emulsion droplets, and found that the micro-explosion strength increases with the ambient temperature. Meng et al.\ \cite{Meng2021DifferentOxygenConcentrations} considered the micro-explosion and combustion of biodiesel-and-ethanol-mixture droplets in high-temperature oxygen environments. They defined a parameter to quantify the micro-explosion intensity and found the environment temperature and the oxygen concentration are important for the micro-explosion and ignition of the droplets. Zhou et al.\ \cite{Zhou2021CombustionCharacteristics} investigated the micro-explosive combustion of biodiesel and DMF (2,5-dimethylfuran) blended fuel droplets. They found that the number and diameter of bubbles in the droplets are the two major factors that determine the intensity of the micro-explosion, and larger bubbles before the rupture results in intense micro-explosion.

In this experimental study of the micro-explosion of emulsion droplets, we find that mist can be generated during the heating of emulsion droplets, and the mist formation is closely related to the micro-explosion process. To explore the mechanism of the mist formation, we perform gas chromatography measurements to test the chemical content of the mist besides high-speed imaging to explore the micro-explosion dynamics. We find the mist formation is due to the vaporization of the emulsion droplet and subsequent condensation of the vapor into many small droplets. Different emulsified fuels are tested in the experiment, and the results show that mist can form for all the emulsified fuels. The mist formation is quantitatively analyzed via digital image processing of high-speed images. The characteristics of the mist in different modes of micro-explosion are investigated, and the effects of the water and surfactant contents in the emulsion droplet are analyzed.

\section{Experimental method}\label{sec:sec2}
\subsection{Preparation of emulsified fuels}\label{sec:sec21}
In this study, n-dodecane, n-hexadecane, and commercial diesel were used as base fuels, and Span 80 (HLB = 4.3) was used as a surfactant to prepare water-in-oil (W/O) emulsified fuels. The surfactant was first added to the oil and mixed by a magnetic stirrer at 3000 rpm for about 5 minutes. After that, deionized distilled water was added to the mixture of oil and surfactant. Then, the magnetic stirrer was used to stir for another 30 minutes at 3000 rpm, so that water was stably dispersed in the form of small water droplets. To avoid the possible influence of emulsion component stratification on the experiments, all experiments were completed within an hour after the preparation of the emulsified oil. The properties of pure liquids used in this study are shown in Table \ref{tab:tab1}. To study the mechanism of the process and the effect of the fuel type, we compared the results of emulsion droplets with different types of oils, i.e., n-dodecane, n-hexadecane, and diesel (in Section \ref{sec:sec31}). To study the influence of the water content in emulsion droplets on the micro-explosion process, the water content was varied from 5\% to 50\% (in Section \ref{sec:sec32}). To study the influence of the surfactant content in the emulsion droplets, the surfactant content was varied from 0.5\% to 4\% (in Section \ref{sec:sec33}).

\begin{table}[]
\caption{Properties of liquids used in this study.}
\label{tab:tab1}
\resizebox{\columnwidth}{!}{%
\begin{tabular}{llllll}
\hline
Liquid       & Density                                                 & Specific heat capacity & Surface tension                                        & Dynamic viscosity                                      & Heat of vaporization                                   \\
             & $\rho$ [kg/m$^3$]                                       & $c_p$ [kJ/(kg$\cdot$K)] & $\sigma$ [mN/m]                                        & $\mu$ [mPa$\cdot$s]                                  & $h_{fg}$ [kJ/kg]                                       \\
\hline
Water        & 998.0 \cite{Jang2021WaterOilEmulsionEmulsionProperties} & 4.18                   & 72.8 \cite{Jang2021WaterOilEmulsionEmulsionProperties} & 0.89 \cite{Jang2021WaterOilEmulsionEmulsionProperties} & 2259 \cite{Jang2021WaterOilEmulsionEmulsionProperties} \\
n-Dodecane   & 744.9 \cite{Zhao2019n-DodecaneDensitySurfaceTension}    & 2.21                   & 23.37 \cite{Zhao2019n-DodecaneDensitySurfaceTension}   & 1.30 \cite{Zhao2019n-DodecaneDensitySurfaceTension}    & 360                                                    \\
n-Hexadecane & 770.3 \cite{Jang2019SurfaceTensionLightAbsorbance}      & 2.21                   & 27.15 \cite{Jang2019SurfaceTensionLightAbsorbance}     & 3.06 \cite{Jang2019SurfaceTensionLightAbsorbance}      & 354                                                    \\
Diesel       & 820 \cite{Antonov2019TwoComponentDropsSuspensions}      & 2.0                    & 27.0 \cite{Antonov2019ImmiscibleFluidsFuelAerosol}     & 2.9 \cite{Antonov2019ImmiscibleFluidsFuelAerosol}      & 210 \cite{Antonov2019TwoComponentDropsSuspensions}     \\ \hline

\end{tabular}%
}
{\small \raggedright Note: The specific heat capacity and the heat of vaporization are from the NIST Chemistry Webbook \cite{NIST}. \par}
\end{table}

\subsection{Experimental details}\label{sec:sec22}
\begin{figure}
  \centering
  \includegraphics[scale=0.6]{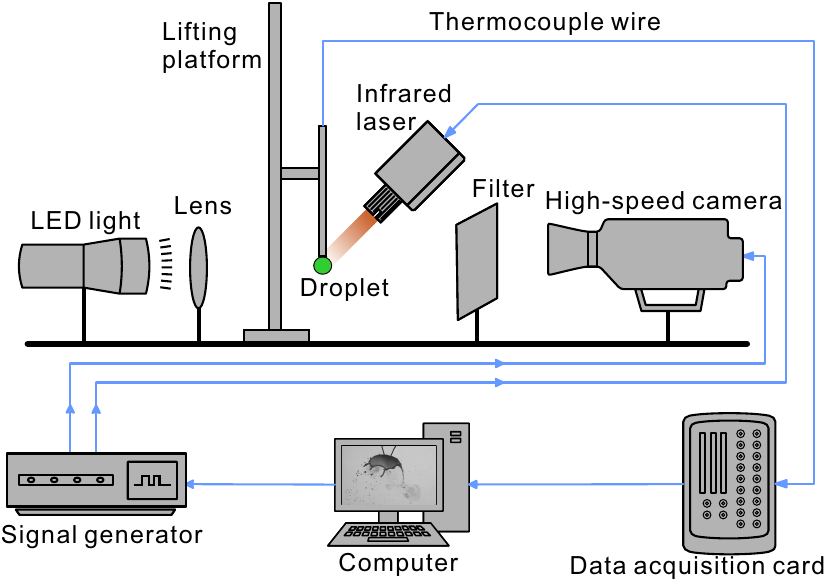}
  \caption{Schematic diagram of the experimental setup.}
  \label{fig:fig01}
\end{figure}

The experimental setup is shown in Figure \ref{fig:fig01}. The emulsion droplet was suspended on a K-type thermocouple junction (wire diameter 120 $\mu$m). The initial droplet diameter is about 1.3--1.5 mm, which is more than 10 times of the thermocouple diameter. The thermocouple was connected to a data acquisition card (National Instruments, NI-USB-6361). The pendant droplet method not only allows us to control the droplet position accurately, but at the same time, allows us to monitor the transient droplet temperature in real time, which is very important for the micro-explosion process. To heat the droplet, a continuous infrared laser (1064 nm, CNI Laser, MIL-H-1064) was used to irradiate the droplet suspended on the thermocouple. Compared with a heating block, both the intensity and the time of the energy input by the laser can be better controlled. The position of the laser spot was adjusted by an infrared light modulator to ensure that the laser spot irradiates the droplet suspended on the thermocouple. In this study, the power of the laser was fixed at the maximum power (about 2W) in the experiment.

In the experiment, a high-speed camera (Photron Fastcam SA1.1) was used to record the dynamics of droplets. The frame rate of the camera was 8000 frames per second (fps), and the image resolution was 15.7 $\mu$m per pixel. A macrolens (LAOWA 100 mm F/2.8D) with a focal length of 100 mm and a magnification of X2 was used. A short exposure time (2 $\mu$s) was used to freeze the fast process in the images. A small aperture (F22) was set in the experiment to obtain enough depth of field to ensure that most secondary droplets could be captured clearly in the image during the breakup. To ensure enough brightness with such a small aperture, a high-power LED light (280 W) scattered by a piece of ground glass was used as the background light source. An optical notch filter (1064 nm) was used to avoid camera damage caused by excessive infrared laser scattering. In the analysis of the results, we performed many repeated experiments to reduce the influence of uncertainty on the analysis. The numbers of repeated experiments used to plot figures are provided in the figure captions or the contexts.

\section{Results and discussion}\label{sec:sec3}
\subsection{Mist formation during the micro-explosion and puffing of emulsion droplets}\label{sec:sec31}

\begin{figure}
  \centering
  \includegraphics[scale=0.7]{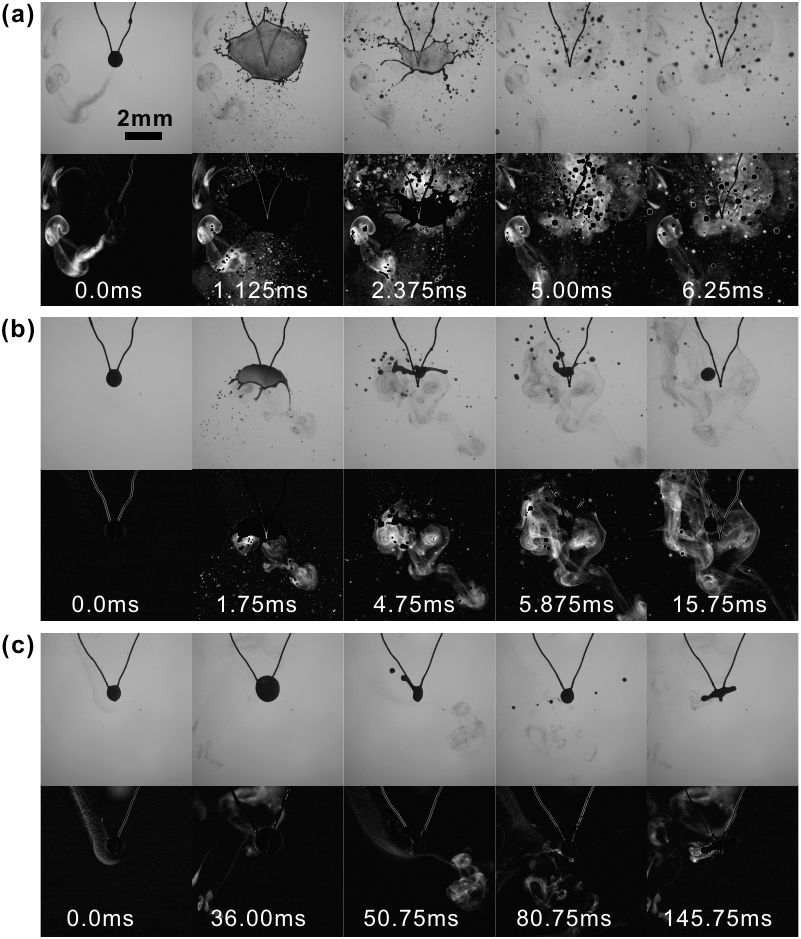}
  \caption{Mist formation during the micro-explosion and puffing of emulsion droplets: (a) intense micro-explosion, (b) weak micro-explosion, and (c) puffing. The oil phase is diesel. The contents of oil, water, and surfactant in the emulsion droplets are 68.5, 30, and 1.5 vol\%, respectively. The first rows are the high-speed images, and the second rows highlight the mist in the images obtained by image processing. Video clips for these processes are available in Supplementary Material (Movies 1-3).}
  \label{fig:fig02}
\end{figure}

\subsubsection{Phenomenon of mist formation}\label{sec:sec311}
In the experiment, we find that mist is generated during micro-explosion and puffing of the emulsion droplets, as shown in Figure \ref{fig:fig02}. The amount of mist strongly depends on the modes of droplet breakup. During the puffing process, several secondary droplets are produced from the main droplet, as shown in Figure \ref{fig:fig02}c. Quickly after a puffing breakup, we can see the formation of mist near the droplet. From the shape evolution of the mist cloud, we can see a sudden current of airflow, which pushes the mist cloud away from the droplet. Therefore, the shape of some mist clouds is like a vortex ring. As the puffing process repeats, the mist is generated again and again.

In the case of intense micro-explosion, the droplet is suddenly stretched into a liquid film, which immediately breaks up, as shown in Figure \ref{fig:fig02}a. Hence, many tiny secondary droplets are produced and ejected. At the moment of the intense micro-explosion, much mist is generated. The micro-explosion can consume all the droplet liquid, which also generates a rapid flow of the nearby air. Hence, after the micro-explosion, the mist cloud is dispersed to the surrounding. When the micro-explosion is weak, the droplet does not break up totally but locally, as shown in Figure \ref{fig:fig02}b. At the moment of the breakup, the mist increases significantly, then the mist is dispersed around, resulting in a substantial decrease in the mist concentration. Compared with the intense micro-explosion, the mist concentration is much lower and the whole process is much slower.

\begin{figure}
  \centering
  \includegraphics[scale=0.7]{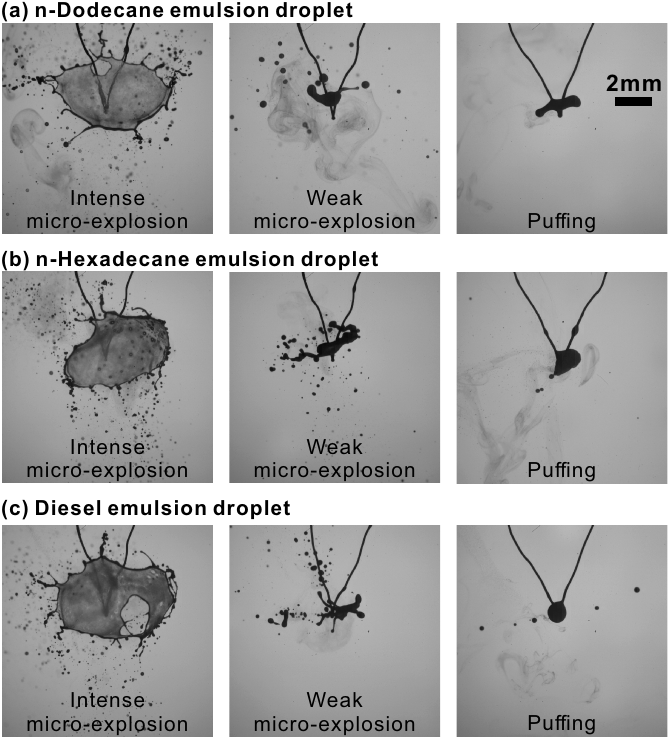}
  \caption{Breakup of emulsion droplets with different oils. (a) n-dodecane emulsion droplet, (b) n-hexadecane emulsion droplet, (c) diesel emulsion droplet. The contents of oil, water, and surfactant in the emulsified oil are 68.5, 30, and 1.5 vol\%, respectively.}
  \label{fig:fig03}
\end{figure}

To confirm that mist formation is not a unique phenomenon of the diesel used in the experiment, we performed more experiments by using different oils. The phenomenon of micro-explosion and mist formation occurs for all the emulsified fuels considered in this study, namely n-dodecane, n-hexadecane, and diesel emulsion droplets, as shown in Figure \ref{fig:fig03}. More results with different contents of water and surfactant will be shown in Sections \ref{sec:sec32} and \ref{sec:sec33}. These results confirm the generality of the mist formation phenomenon during the puffing and micro-explosion of emulsion droplets.

Even though several studies mentioned mist or aerosol during the micro-explosion of emulsion droplets, they are different processes. For example, Antonov et al.\ \cite{Antonov2019TwoComponentDropsSuspensions} reported that at the moment of the droplet micro-explosion, the intense disintegration of the main droplets produces fine aerosol. Gao et al.\ \cite{Gao2022PolymerSurfactant} found for the micro-explosion of emulsion droplets, the instantaneous evaporation of the water phase causes the droplet to burst into mist. Jang et al.\ \cite{Jang2021WaterOilEmulsionEmulsionProperties} found that for an emulsion droplet exposed to a high-power laser pulse ($\geq$ 11.5 mJ), mist is ejected from the main droplet. Therefore, in these studies, aerosol/mist is produced at the moment of the droplet micro-explosion from the direct breakup of the main droplet. In contrast, in our experiment, the mist is formed even before the micro-explosion, as shown in Figure \ref{fig:fig02}a. In addition, in our experiment, the mist is formed during the process of puffing, whereas aerosol/mist in previous studies occurs only for intense micro-explosion.

\subsubsection{Mechanism of mist formation}\label{sec:sec312}
It should be noted that the mist is not produced by the direct breakup of the emulsion droplets, even though much mist is produced during the breakup. This can be confirmed by the high-speed images of the mist formation process. The direct production of mist (i.e., tiny droplets) on the surface of the droplet would require violent movement and deformation of the interface, which does not exist in the mist formation process. Particularly, the direct formation of such tiny droplets is impossible during the puffing process, during which the interface movement is very mild. We can also see that the mist is formed mainly slightly after the instant of the breakup and a small amount even before the instant of the breakup, not only at the exact moment of the breakup. In addition, some mist is formed at a certain distance from the emulsion droplet. These features of mist generation indicate that the mist is not directly produced from the droplet.

\begin{figure}
  \centering
  \includegraphics[scale=0.38]{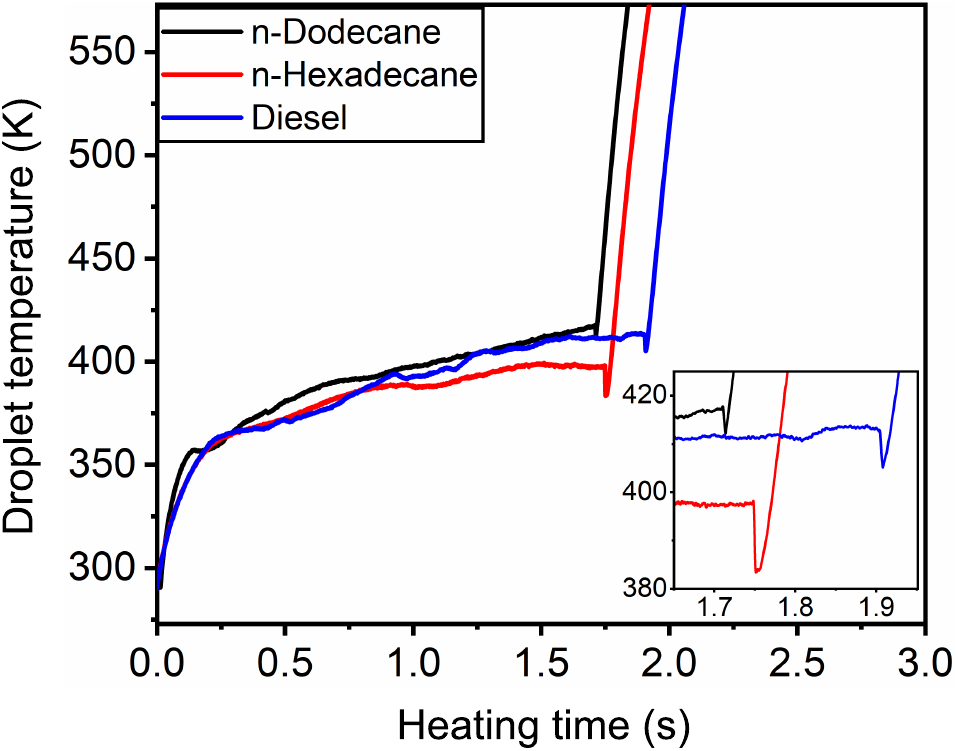}
  \caption{Temperature variation of emulsion droplets corresponding to the intense micro-explosion in Figure \ref{fig:fig03}.}
  \label{fig:fig04}
\end{figure}

Based on the experimental observation, we hypothesize that the formation of the mist is due to the condensation of the vapor. At the moment of droplet breakup, either puffing or micro-explosion, a large amount of vapor is produced, which can be confirmed by the temperature reduction at the moment of the breakup, as shown in Figure \ref{fig:fig04}. The temperature reduction is due to the sudden vaporization, which absorbs a large amount of latent heat of vaporization \cite{Shen2019AlternativeFuel}. Similar temperature changes were found in many studied of micro-explosion process \cite{Mura2012WaterMetastableState, Shen2019AlternativeFuel, Watanabe2010BreakupCharacteristics}. For the three intense micro-explosion processes in Figure \ref{fig:fig03}, the temperature reduction at the moment of the breakup is, respectively, 5.1, 14.2, 7.3 K for n-dodecane, n-hexadecane, and diesel emulsion droplets, as shown in Figure \ref{fig:fig04}. The breakup also induces a rapid flow of air, which pushes the vapor away from the droplet. As the nearby air temperature is low, the vapor can quickly condense, forming many tiny droplets, i.e., mist.

Another explanation of the phenomenon might be the formation of solid particles via chemical reaction under heating. But this is less possible, because in the experiment, the temperature is always below 923 K, and we did not observe any flame in the process. To further test this hypothesis, we checked the chemical content of the mist. If it is particles produced after combustion or pyrolysis, some new species would exist in the air. In contrast, if the mist is produced via vaporization and condensation (i.e., a purely physical process), the chemical content of the vapor/gas, besides air, should be the same as the oil used in the experiment.

To check the chemical content of the mist, we performed gas chromatographic measurements. Micro gas chromatography (Micro GC Fusion) was used to detect the mist produced from the emulsion droplets of different compositions. Gas chromatograph uses the difference in boiling point, polarity, and adsorption properties to separate mixtures. Firstly, the vaporized gas was brought into the chromatographic column of the gas chromatograph by the inert gas. Due to the flow of the inert gas, the gas components are adsorbed during the movement. As a result, the different components are separated after flowing through the chromatographic column. When the components flow out of the chromatographic column, they immediately enter the detector, and then the concentration of the sample components can be transformed into electrical signals, showing chromatographic peaks.

\begin{figure}
  \centering
  \includegraphics[width=\columnwidth]{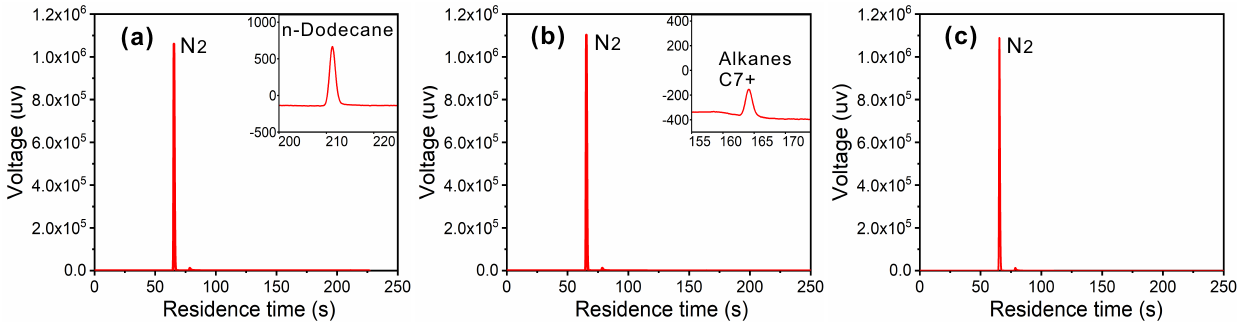}
  \caption{Gas chromatography analysis of the mist formed during heating of emulsion droplets: (a) n-dodecane emulsion droplet, (b) diesel emulsion droplet, (c) n-hexadecane emulsion droplet. The curves for O$_2$ are from another chromatographic column and are not shown here. The contents of oil, water, and surfactant in the emulsified oil are 68.5, 30, and 1.5 vol\%, respectively.}
  \label{fig:fig05}
\end{figure}

The gas chromatography result for the mist produced by the droplet explosion and puffing process of n-dodecane emulsified oil is shown in Figure \ref{fig:fig05}a. It can be found that there is n-dodecane in addition to N$_2$ and O$_2$ (air components). Similarly, for the chromatographic result of diesel emulsified oil shown in Figure \ref{fig:fig05}b, there is a peak at about 165 s corresponding to long-chain alkanes (C7+) in addition to N$_2$ and O$_2$, because diesel is a mixture of many alkanes. For the chromatographic result of n-hexadecane emulsified oil in Figure \ref{fig:fig05}c, only N$_2$ and O$_2$ were detected, but n-hexadecane was not detected. The reason is that the injection temperature of gas chromatography is about 493 K, which is below the boiling point of n-hexadecane (559.79 K). Hence, the n-hexadecane did not vaporize, but existed in liquid form, so n-hexadecane could not be detected. The gas chromatography confirms that there is no low-carbon substance during the mist formation, such as carbon monoxide, C2--C4 alkenes, and acetylene, which are calibrated but not detected. Therefore, we can exclude the combustion or pyrolysis hypothesis, and confirm that the process of mist formation is a physical process.

Therefore, we can confirm that the mechanism of the mist formation is due to vaporization and condensation. Under the heating of the droplet, water or even some oil in the emulsion droplet vaporizes to produce vapor. At the moment of puffing and micro-explosion, a large amount of liquid suddenly changes into vapor. As the vapor is dispersed to the surrounding, the vapor condenses into small droplets due to the decrease in the environment temperature. Of course, the mist formation is a very complex process, and further investigation of the mechanism is required.

\subsubsection{Characteristics of mist }\label{sec:sec313}
\begin{figure}
  \centering
  \includegraphics[scale=0.52]{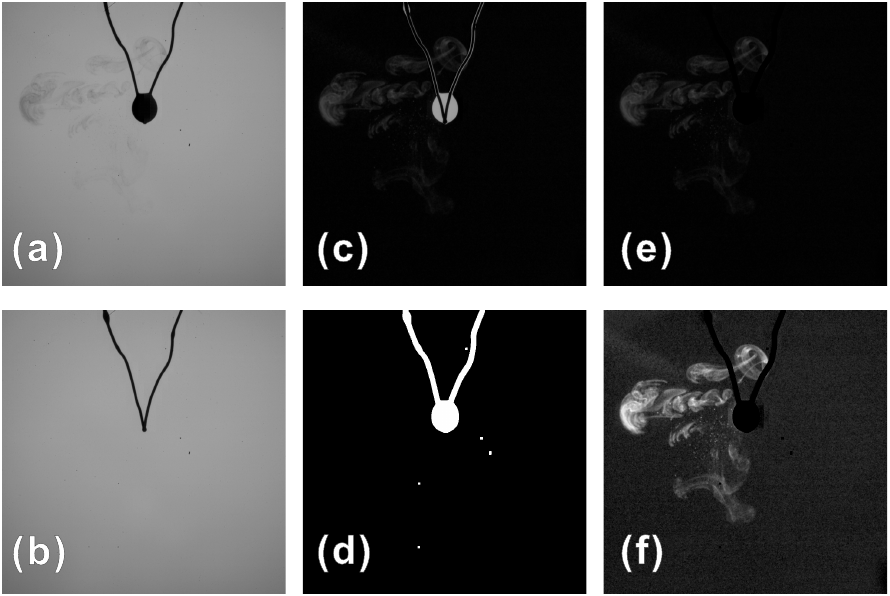}
  \caption{Schematic diagram of measurements of mist concentration in image processing. (a) Original image. (b) Background image before the experiment without the emulsion droplet, secondary droplets, or mist. (c) Difference between the original image with the background image. (d) Mask image obtained based on the brightness of the original image and the background image. (e) Mist image. (f) Mist image with enhanced contrast.}
  \label{fig:fig06}
\end{figure}

To quantify the mist concentration, we performed digital image analysis using a customized Matlab program to measure the concentration of the mist from high-speed images. The schematic diagram of the mist characteristic measurement in the Matlab program is shown in Figure \ref{fig:fig06}. First, the raw image (Figure \ref{fig:fig06}a) was filtered by a wiener filter to remove noise. Then, by subtracting an image before the start of the experiment (without droplet or mist, Figure \ref{fig:fig06}b), the background was removed, as shown in Figure \ref{fig:fig06}c. A mask was obtained based on the image brightness as shown in Figure \ref{fig:fig06}d, including the thermocouple, the emulsion droplet, and secondary droplets. After that, the mist can be identified by applying the mask to the image without the background, as shown in Figure \ref{fig:fig06}e. The mist can be seen clearly by increasing the image contrast, as shown in Figure \ref{fig:fig06}f. Then, the concentration of the mist was characterized by the cumulative gray value of the mist image (Figure \ref{fig:fig06}e). The higher the accumulated gray value in the image is, the higher the mist concentration is.

\begin{figure}
  \centering
  \includegraphics[width=\columnwidth]{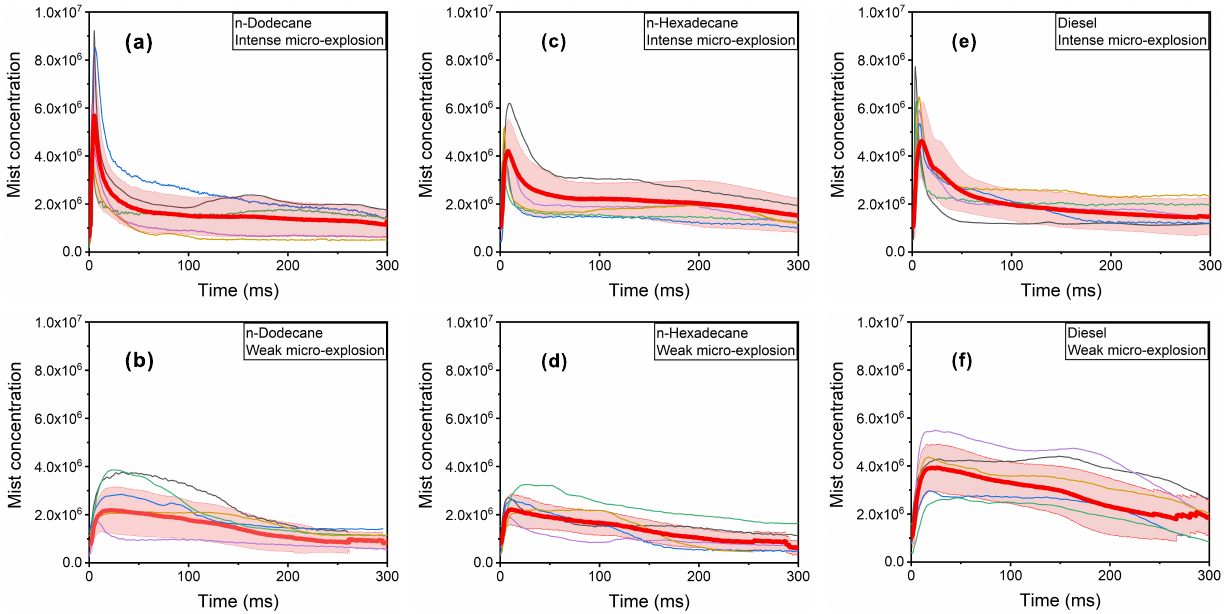}
  \caption{Time variation of mist concentration during the micro-explosion. (a,c,e) Intense micro-explosion; (b,d,f) Weak micro-explosion. The contents of oil, water, and surfactant in the emulsified oil are 68.5, 30, and 1.5 vol\%, respectively. Each thin line represents the result of a typical case, and five cases are shown in each panel. The thick red lines and their error bands represent the average and the standard deviation of all experimental cases. The numbers of experimental cases from (a) to (f) are respectively 18, 16, 12, 22, 18, and 17.}
  \label{fig:fig07}
\end{figure}

The time evolution of the mist concentration for intense and weak micro-explosion is remarkably different, as shown in Figure \ref{fig:fig07}. For intense micro-explosion, the peak is high and thin, indicating a sudden formation of the mist during the intense micro-explosion. In contrast, for weak micro-explosion, the peak is much lower and wider, indicating the mist is formed for a longer time, but with weaker strength. During the intense micro-explosion, a large amount of water was vaporized inside the droplet, so a large amount of heat was absorbed. After the condensation of water vapor, dense mist was produced. Therefore, the curve had a sharp rise. After the micro-explosion, the curve has a sharp decrease as the mist disperses around. The intense micro-explosion also produces a strong air current which accelerates the dispersion of the mist. It takes only about 50 ms for the curve to reach a steady state. In the case of weak micro-explosion, only a small amount of water phase vaporized. Hence, only a small amount of gas was sprayed from a certain position on the surface of the oil droplet during micro-explosion, resulting in the explosion of only part of the emulsion droplet. Therefore, the mist was less than that generated by the intense micro-explosion, and the magnitude of the peak is much lower. But it takes a longer time (about 200--300 ms) for the curve to become stable.

To quantitatively describe the mist formation in the micro-explosion process, we introduced two variables. To quantify the concentration of the mist, we used $\Delta P$ defined as
\begin{equation}\label{eq:eq01}
  \Delta P={{P}_{\max }}-P
\end{equation}
where $P _ {\max}$ and $P$ are the maximum and the final stable values of the mist concentration, respectively, which could be obtained from the mist concentration curve. To quantify the duration of the mist, we used $T_{80}$, which is the duration when the mist concentration is above $0.8 P_{\max}$ in the mist concentration curve. The larger the concentration difference $\Delta P$ is and the shorter the duration $T_{80}$ is, the more intense the micro-explosion process is. Therefore, the two parameters, the concentration $\Delta P$ and the duration $T_{80}$, are important to describe the mist formation quantitatively.

These two parameters, $T_{80}$ and $\Delta P$, were used to characterize the change of the micro-explosion mode. In the intense micro-explosion mode, a large amount of mist was generated after the droplet intensely breaks up. After the micro-explosion was completed, the mist quickly disperses around. Therefore, in the intense micro-explosion mode, the duration $T_{80}$ is very small, and the concentration $\Delta P$ is very large. In contrast, in the weak micro-explosion mode, the mist concentration $\Delta P$ is much smaller than that under the intense micro-explosion mode. After the micro-explosion, the mist disperses around, but the dispersion speed is very slow, hence the duration $T_{80}$ increases significantly.

\begin{figure}
  \centering
  \includegraphics[scale=0.45]{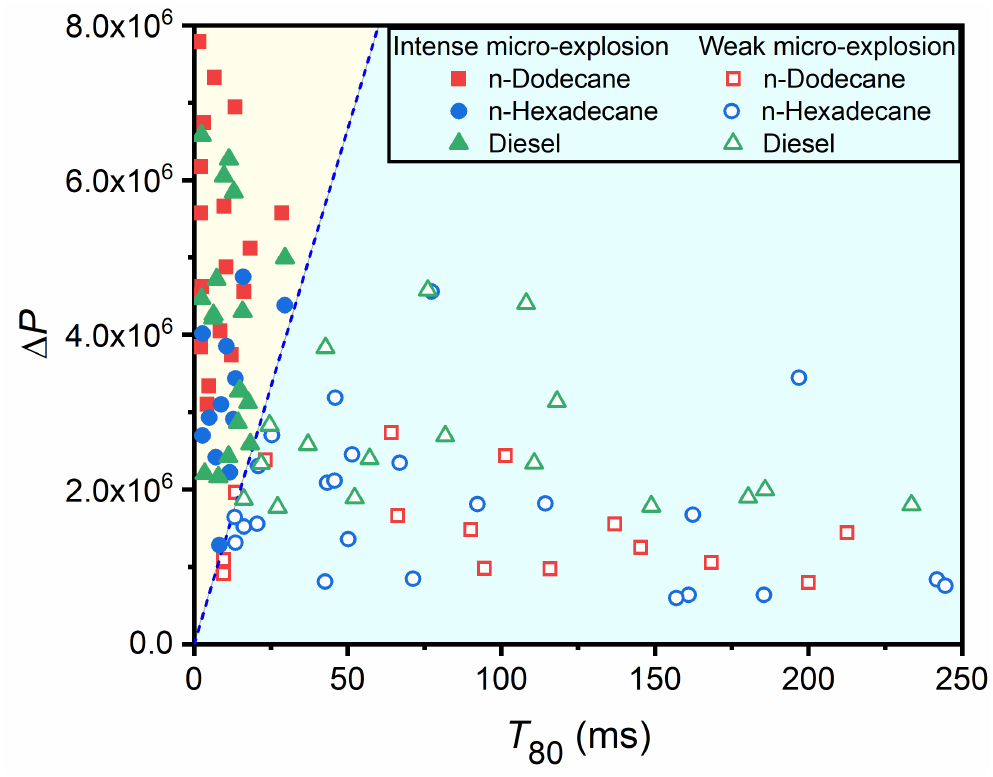}
  \caption{Mist characteristics for intense and weak micro-explosion. The two parameters,  $T_{80}$ and $\Delta P$, are the duration and the concentration of the mist.}
  \label{fig:fig08}
\end{figure}

Then we plot the concentration $\Delta P$ and the duration $T_{80}$ in a map for all micro-explosion processes, as shown in Figure \ref{fig:fig08}. We can see the intense micro-explosion is mainly in the upper left part of the map, while the weak micro-explosion mode is mainly in the lower right part. With these two parameters, the duration $T_{80}$ and the concentration $\Delta P$, we can distinguish the micro-explosion modes by using a straight line,
\begin{equation}\label{eq:eq02}
  \Delta P=1.3 \times 10 ^5 \cdot T_{80}
\end{equation}
as shown in Figure \ref{fig:fig08}.

\begin{figure}
  \centering
  \includegraphics[scale=0.42]{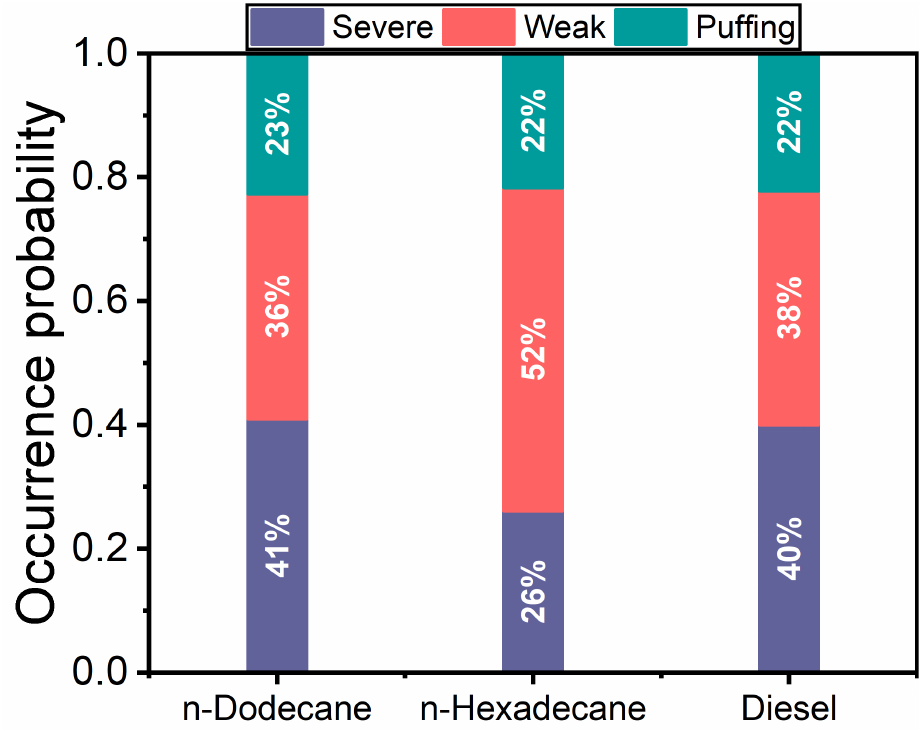}
  \caption{Effect of different compositions in emulsified fuels on the breakup mode. The probability was obtained by conducting 30 groups of experiments using the emulsion droplets of each composition.}
  \label{fig:fig09}
\end{figure}

The composition of the emulsion droplet also affects the micro-explosion. For different fuels, the occurrence probabilities of puffing and weak/intense micro-explosion are presented in Figure \ref{fig:fig09}. It can be found that the type of oil influences the micro-explosion mode. n-Hexadecane has a lower probability of intense micro-explosion but a higher probability of weak micro-explosion. The difference can be attributed to the properties of the oil in the emulsion droplet, particularly the surface tension and the viscosity. It has been shown that the surface tension and the viscosity of the oil in emulsion droplets have an important influence on the micro-explosion \cite{Antonov2019TwoComponentDropsSuspensions}. When the surface tension and the viscosity are large, the droplets are more difficult to deform and break up. Hence the intensity of the micro-explosion is relatively weak. With the decrease of the surface tension and the viscosity, less energy is needed for the deformation and fragmentation of the droplet. Therefore, with higher surface tension and viscosity (as shown in Table \ref{tab:tab1}), n-hexadecane has a much lower probability of intense micro-explosion than n-dodecane and diesel.

\subsubsection{Characteristics of secondary droplets}\label{sec:sec314}

\begin{figure}
  \centering
  \includegraphics[width=\columnwidth]{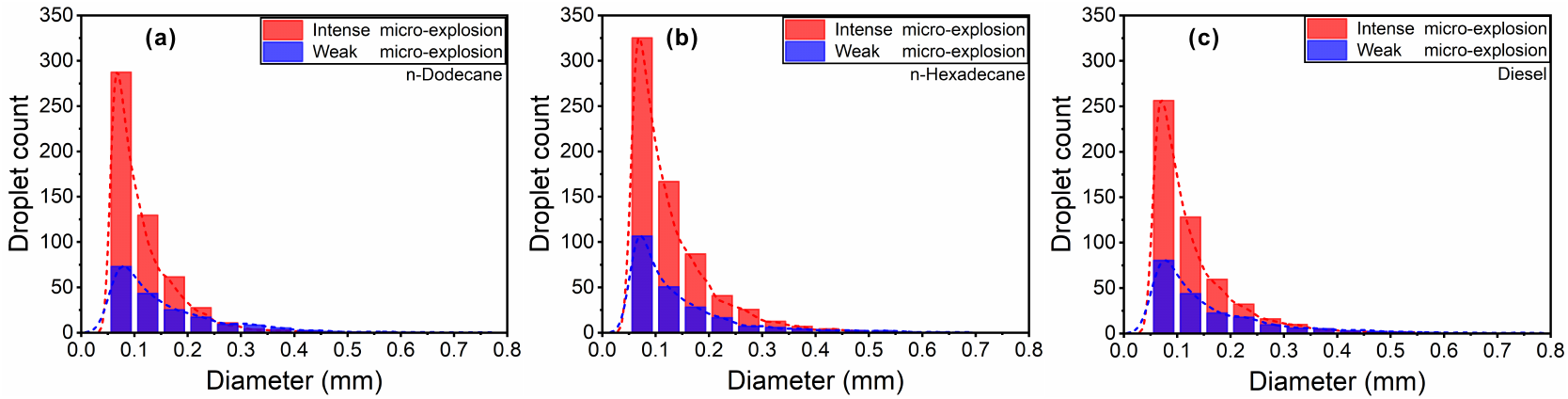}
  \caption{Size distribution of the secondary droplets produced after the micro-explosion: (a) n-dodecane emulsion droplets; (b) n-hexadecane emulsion droplets; (c) diesel emulsion droplets. The contents of oil, water, and surfactant in the emulsified oil are 68.5, 30, and 1.5 vol\%, respectively. }
  \label{fig:fig10}
\end{figure}

The size distribution of the secondary droplets during micro-explosion was obtained via image processing, as shown in Figure \ref{fig:fig10}. The number and the diameter of secondary droplets were measured from high-speed images using a customized Matlab program. For each component of emulsified oil droplets, ten groups of experiments under intense micro-explosion and weak micro-explosion modes were selected, and the average for the ten groups of micro-explosion was used to plot the size distribution of secondary droplets.

As shown in Figure \ref{fig:fig10}, almost all secondary droplets are smaller than 0.5 mm in diameter, and the highest probability of the secondary droplets is below 0.1 mm. Similar size distribution of secondary droplets was also obtained by Pavel et al.\ \cite{Strizhak2021ChildDropletsSchlierenPhotography}, who studied the micro-explosion of emulsion droplets at different temperatures, and found that the maximum probability distribution of secondary droplet size is below 0.1 mm. Comparing intense and weak micro-explosion, we can see that the total number of secondary droplets produced after intense micro-explosion is several times more than that for weak micro-explosion, indicating that the intense micro-explosion has a better capability in atomization. This is consistent with many previous studies\ \cite{Strizhak2021ChildDropletsSchlierenPhotography, Rosli2021SuspendedDroplet, Lyu2021MutualSolubilityDifferentials}, which found that a high micro-explosion intensity promotes intense breakup of emulsion droplets, thereby resulting in numerous secondary droplets. Taking n-dodecane for example (see Figure \ref{fig:fig10}a), the percentage of the secondary droplet distribution in the range of 0--0.1 mm is about 54\% and 38\% for intense and weak micro-explosion, respectively. For the range of 0--0.2 mm, the percentages are 90\% and 74\% for intense and weak micro-explosion, respectively. Hence, the secondary droplets produced after weak micro-explosion are much larger than that for intense micro-explosion.

The size distribution of the secondary droplet is because of the process of the micro-explosion, whose strength directly affects the stretching of the droplet during the micro-explosion process. In the case of intense micro-explosion, the whole droplet is stretched into a film, which spreads intensely and completely breaks into many secondary droplets. Hence, the size of the secondary droplets is relatively small. In the case of weak micro-explosion, only part of the droplet explodes. After the droplet is stretched to a certain extent, it retracts to form a droplet, which is much larger than secondary droplets. Therefore, the number of secondary droplets is much less than that during intense micro-explosion, and the size of the secondary droplets is much larger.

\subsection{Effects of water contents}\label{sec:sec32}

\begin{figure}
  \centering
  \includegraphics[scale=0.6]{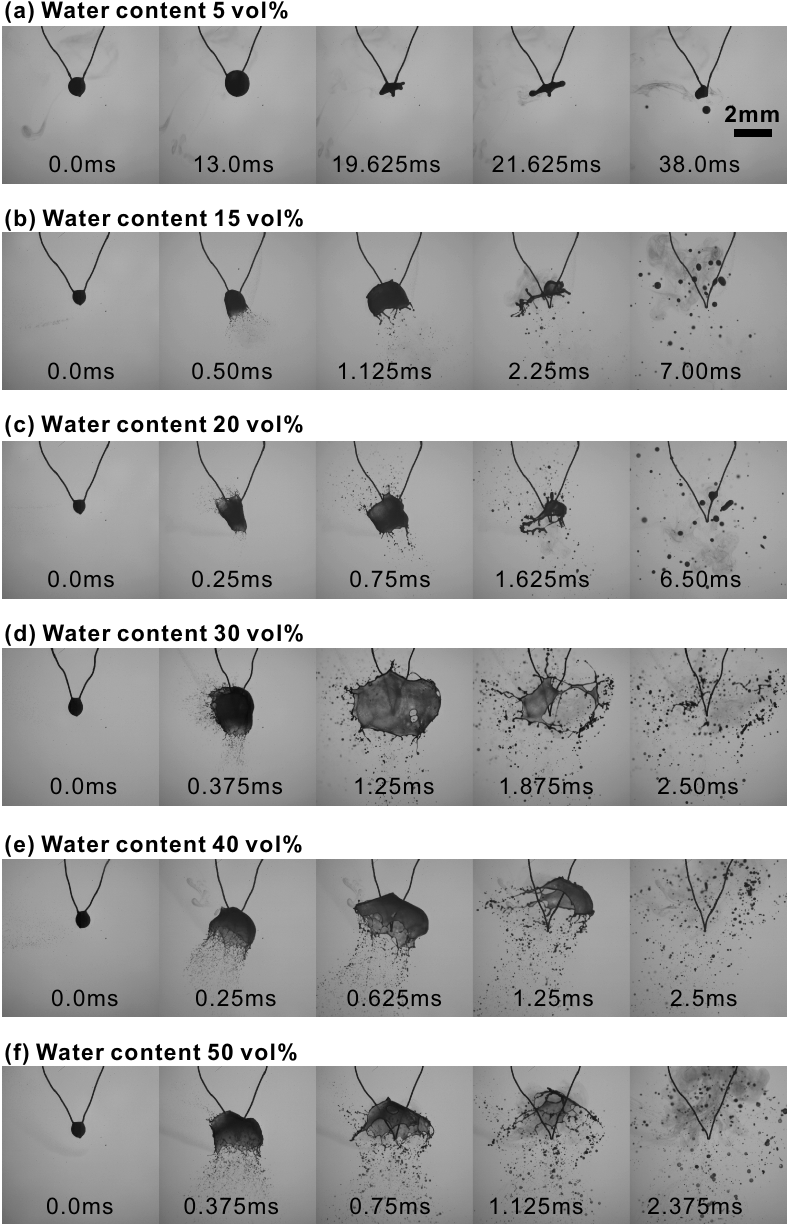}
  \caption{Breakup of emulsion droplets with different water contents. The oil phase is diesel, the surfactant content in the emulsified oil is 1.5 vol\%, and the water content change from 5--50\%.}
  \label{fig:fig11}
\end{figure}

The effect of the water content in emulsion droplets on the micro-explosion characteristics is shown in Figure \ref{fig:fig11}. When the water content of the emulsified oil is 5 vol\%, no micro-explosion is observed in the heating process, but an obvious puffing phenomenon occurred as shown in Figure \ref{fig:fig11}a. The puffing process occurs repeatedly until the water in the emulsion droplets is completely evaporated. When the water content increases to 15--20 vol\%, weak micro-explosion occurs during the heating process, as shown in Figures \ref{fig:fig11}b,c. When the water content further increases to 30 vol\%, intense micro-explosion occurs, as shown in Figure \ref{fig:fig11}d. When the water content continues to increase beyond 30 vol\%, the droplet breakup process is similar to that of 30 vol\%, as shown in Figures \ref{fig:fig11}e,f. Overall, the water content in emulsified oil droplets enhances the breakup strength. With the increase of water content, the breakup strength increases, and the micro-explosion mode gradually changes from puffing to weak micro-explosion and then to intense micro-explosion. This is consistent with previous studies\ \cite{Califano2014SingleDropletStability, MeloEspinosa2018MicroChannelEmulsifier, Cen2019SputteringEmulsionFuel}. For example, Valeria et al.\ \cite{Califano2014SingleDropletStability} found that the micro-explosion at 30\% water content is stronger than that at 10\% water content, and attributed the result to the aggregation of water droplets and the simultaneous vaporization of a large amount of water.

\begin{figure}
  \centering
  \includegraphics[scale=0.45]{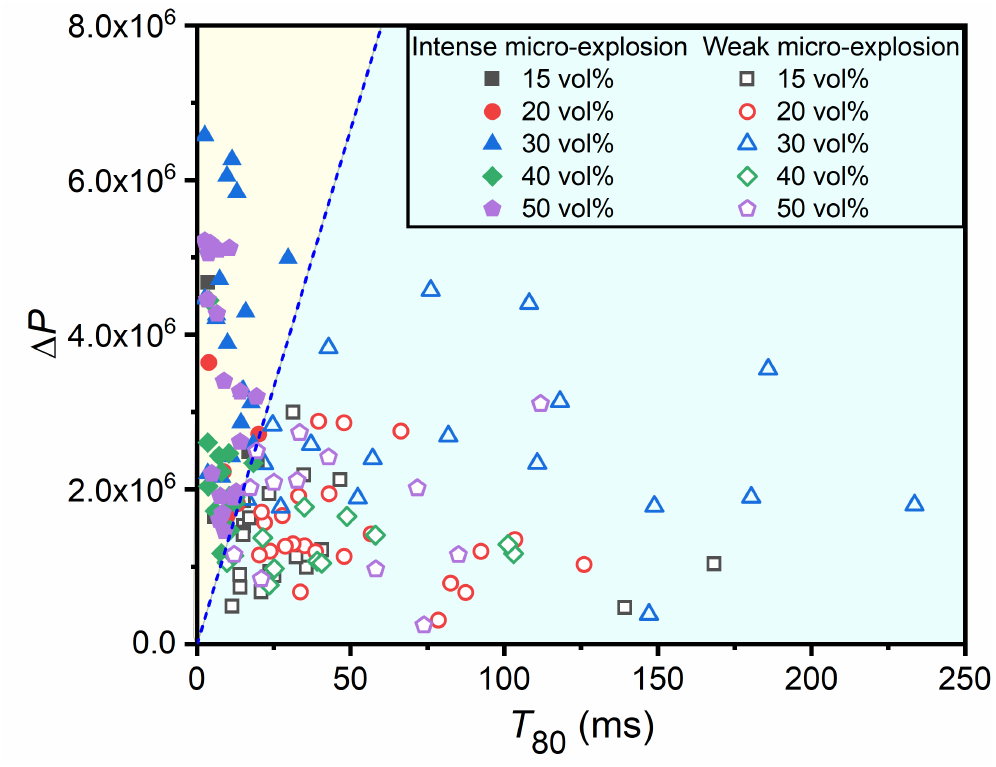}
  \caption{Map of micro-explosion for emulsion droplets of different water contents with the duration ($T_{80}$) and concentration ($\Delta P$) of the mist.}
  \label{fig:fig12}
\end{figure}

A map of the duration ($T_{80}$) and the concentration ($\Delta P$) of the mist for the micro-explosion of emulsion droplets with different water contents are shown in Figure \ref{fig:fig12}. The experiment of each water content was repeated 30 times to study the probabilities of the outcomes. The mist concentration curve was extracted from the experimental images. The micro-explosion of emulsion droplets with different water content is characterized by the duration $T_{80}$ and the concentration $\Delta P$. Then, Figure \ref{fig:fig12} is plotted using the two parameters, and Eq.\ (\ref{eq:eq02}) was used to distinguish the micro-explosion modes. For the intense micro-explosion, the duration $T_{80}$ is very small, and the concentration $\Delta P$ is relatively large. For the weak micro-explosion, the duration $T_{80}$ increases significantly, and the concentration $\Delta P$ is small.

\begin{figure}
  \centering
  \includegraphics[scale=0.42]{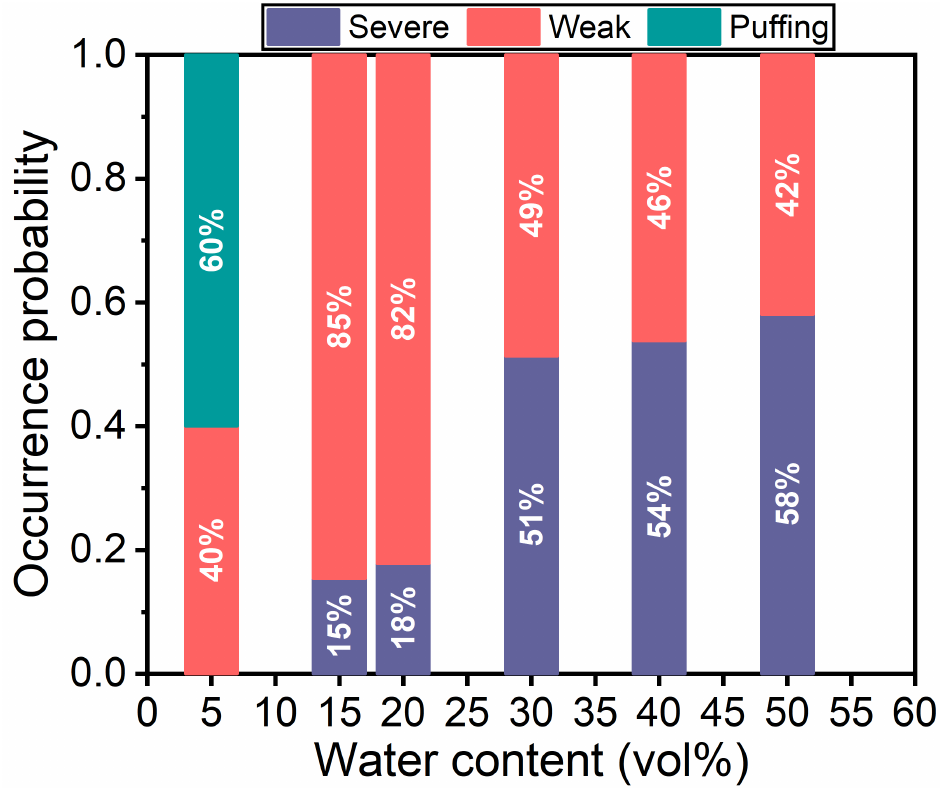}
  \caption{Effect of water content on the breakup mode. The probability was obtained by conducting 30 groups of experiments using the emulsion droplets of each composition.}
  \label{fig:fig13}
\end{figure}

The influence of the water content on the micro-explosion mode is shown in Figure \ref{fig:fig13}. The percentage of intense and weak micro-explosion of emulsion droplets with different water contents was counted according to Figure \ref{fig:fig12}. It can be found that the influence of the water content on the micro-explosion mode is very significant. When the water content is very small (i.e., 5 vol\%), there is a small amount of vaporization, which is not enough to cause intense micro-explosion, resulting in puffing and weak micro-explosion. When the water content is increased to 15--20 vol\%, most droplets have weak micro-explosion due to more vaporization. When the water content continues to increase to 30 vol\%, many droplets have intense micro-explosion due to a large amount of vaporization. However, when the water content increases further, the percentage of intense micro-explosion is not greatly improved. This is because when the water content is too large, the water evaporation is basically maintained at about 20 vol\% \cite{Shen2019AlternativeFuel}. Even if there is more water in the oil droplets, it cannot be vaporized in such a short interval of micro-explosion to improve the breakup strength. Moreover, the excessive water will increase the heating time before the occurrence of the micro-explosion. In summary, the water content in emulsified oil can influence the droplet breakup mode and strength. Maintaining the water content at about 30 vol\% is most conducive to the micro-explosion of emulsion droplets, which significantly increases the percentage of intense micro-explosion.

\subsection{Effects of surfactant contents}\label{sec:sec33}

\begin{figure}
  \centering
  \includegraphics[scale=0.6]{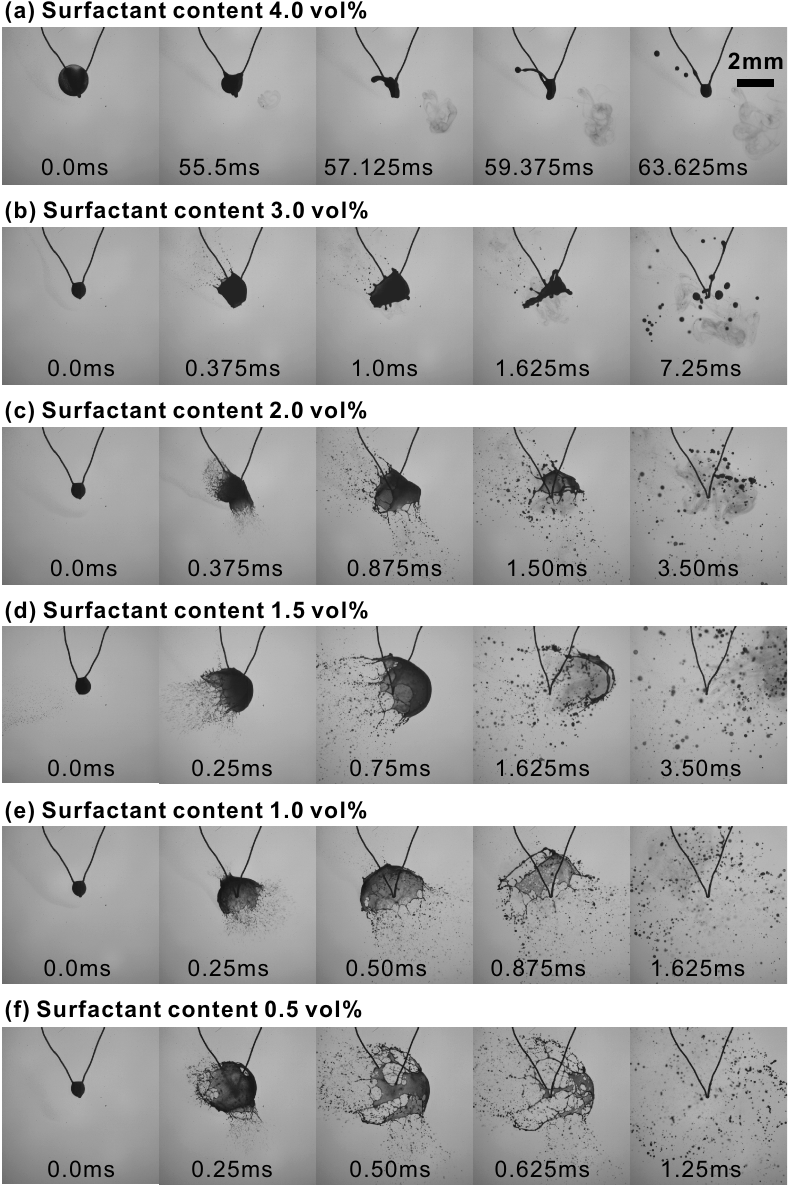}
  \caption{Breakup characteristics of emulsion droplets with different surfactant contents. The oil phase is diesel, the content of water is 30 vol\%, and the surfactant content changes from 0.5--4.0\%.}
  \label{fig:fig14}
\end{figure}

The influence of the surfactant content in the emulsion droplets on the micro-explosion is shown in Figure \ref{fig:fig14}. When the surfactant content is 4 vol\%, no micro-explosion phenomenon is observed in the heating process, but puffing occurs, as shown in Figure \ref{fig:fig14}a. When the surfactant content decreases to 3 vol\%, micro-explosion occurs, as shown in Figure \ref{fig:fig14}b. When the surfactant content is reduced to 2 and 1.5 vol\%, the micro-explosion becomes more obvious, as shown in Figure \ref{fig:fig14}c,d. When the surfactant content continues to decrease to 1 and 0.5 vol\%, as shown in Figures \ref{fig:fig14}e,f, the micro-explosion becomes very intense. In this process, small secondary droplets move very quickly out of the shooting area of the camera, and the whole micro-explosion process is completed within approximately 1--2 ms. Although the droplet had a violent micro-explosion, a large amount of mist was not found in the images. This is because when the surfactant content is particularly small, a large amount of water vaporized, which induces a very rapid gas current and quickly blows the secondary droplet away. Therefore, the mist is rapidly dispersed out of the shooting area (see Movie 4 in the supplementary material). Therefore, the mist concentration in the images is low even though the micro-explosion is intense. Overall, with the increase of surfactant content, the micro-explosion of emulsion droplets transits from intense micro-explosion to weak micro-explosion and then to puffing. This is also consistent with previous studies of the surfactant effect\ \cite{Califano2014SingleDropletStability, MeloEspinosa2018MicroChannelEmulsifier, Suzuki2011AggregationSecondaryAtomization}. For example, Eliezer et al.\ \cite{MeloEspinosa2018MicroChannelEmulsifier} found that an increase in surfactant content led to smaller water droplets and better stability of the dispersed system.

\begin{figure}
  \centering
  \includegraphics[scale=0.45]{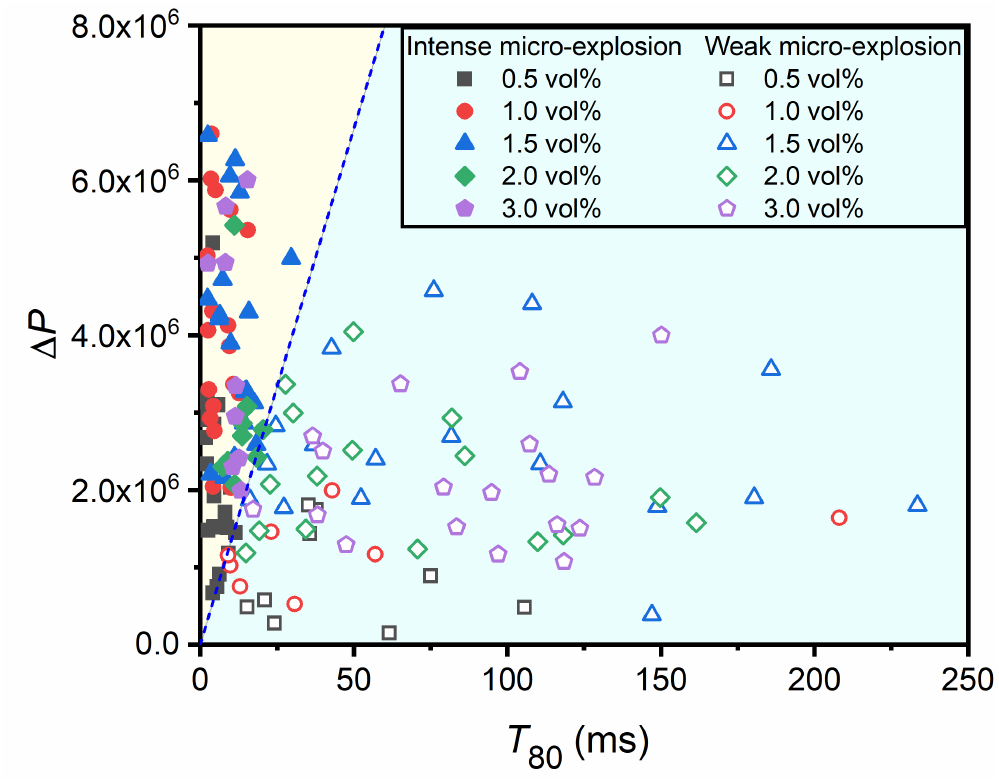}
  \caption{Map of micro-explosion regimes for emulsion droplets of different surfactant contents with the duration ($T_{80}$) and the concentration ($\Delta P$) of the mist.}
  \label{fig:fig15}
\end{figure}

\begin{figure}
  \centering
  \includegraphics[scale=0.42]{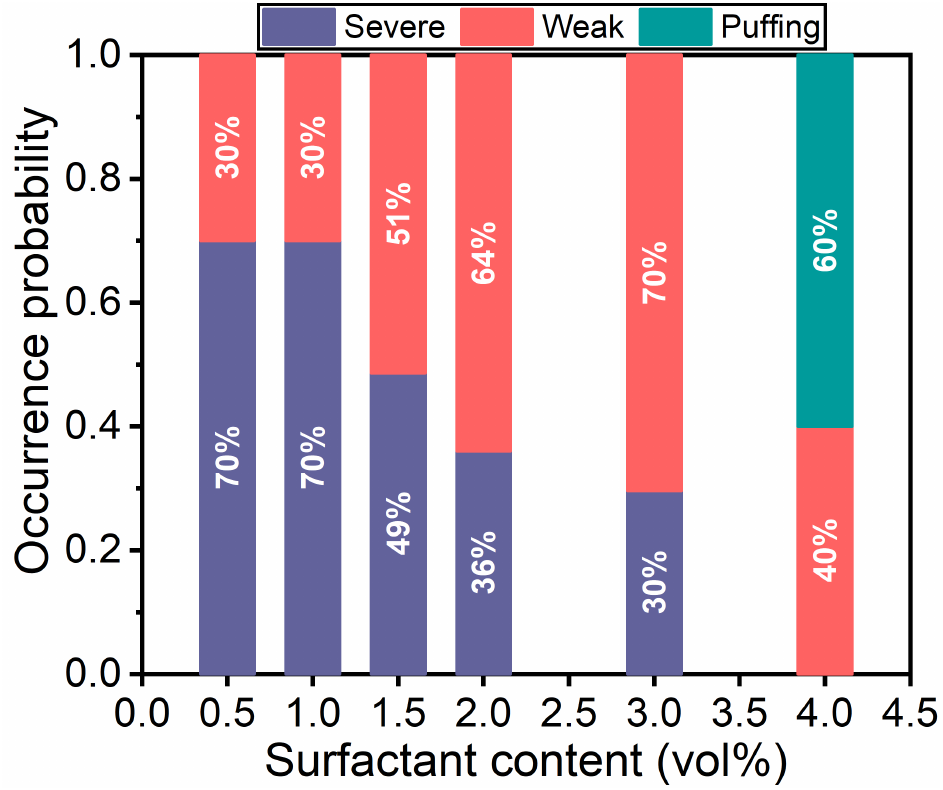}
  \caption{Effect of surfactant content on the breakup mode. The probability was obtained by conducting 30 groups of experiments using the emulsion droplets of each composition.}
  \label{fig:fig16}
\end{figure}

A map of the duration ($T_{80}$) and the concentration ($\Delta P$) of the mist for the micro-explosion of emulsion droplets with different surfactant contents are shown in Figure \ref{fig:fig15}. It is found that the micro-explosion mode can be well distinguished by Eq.\ (\ref{eq:eq02}). The influence of the surfactant content on the micro-explosion mode is shown in Figure \ref{fig:fig16}. The probability of intense micro-explosion was counted according to Figure \ref{fig:fig15}. It can be found that the influence of the surfactant content on the micro-explosion mode is very significant.

The effect of surfactant content on the mode of micro-explosion can be attributed to the surfactant deactivation at high temperature. When the surfactant content is low (i.e., 0.5 and 1 vol\%). During the heating process, the surfactant molecules at the oil-water interface are deactivated under high-temperature conditions \cite{Shen2020DropletBreakup}. Then the coalescence of the water droplet reduces the water-oil interface area which serves as the heterogeneous nucleation sites for vaporization and micro-explosion \cite{Shen2020DropletBreakup}. Therefore, the droplet temperature further increases during the heating process until intense micro-explosion occurs due to the large superheating. In contrast, when the surfactant content is high (i.e., 4 vol\%), the oil-water surface tension is low and can always stabilize the water droplet during the heat process. The larger oil-water interfacial area creates more heterogeneous nucleation sites, hence puffing can occur easily. By periodic puffing to release energy, intense micro-explosion is avoided. As a result, with only a small amount of vaporization during the puffing process, the mist concentration is low, and the period is long. In summary, the surfactant content affects the surfactant deactivation temperature in the heating process, thus further affecting the micro-explosion mode and mist formation.

\section{Conclusions}\label{sec:sec4}
In this study, we find that mist is generated during puffing and micro-explosion of the emulsion droplets. The mechanism of the mist formation is due to the vaporization and condensation. Under the heating of the droplet, water or even some oil in the emulsion droplet vaporizes particularly at the moment of puffing/micro-explosion. Then, the vapor condenses into small droplets due to the temperature decrease. The evolution of the mist concentration is highly correlated with the micro-explosion mode. For intense micro-explosion, the mist concentration curve shows a peak high and thin, indicating a sudden formation of mist during the micro-explosion and a quick dispersion after the micro-explosion. In contrast, the mist concentration peak for weak micro-explosion is lower and wider. Based on these features, two parameters are proposed to characterize the mist during the micro-explosion, i.e., the concentration and the duration, which can be used to distinguish the micro-explosion modes. The effects of the water and surfactant contents in the emulsion droplets are studied, and the mists are used to characterize the micro-explosion. Increasing the water content can promote the vaporization of the water phase, increase the strength of micro-explosion, and result in a large amount of mist. Increasing the surfactant content can improve the stability of the emulsion droplet, reduce the probability of intense micro-explosion, and hence reduce the mist concentration.

The mist formation during puffing and micro-explosion of the emulsion droplets is a complex process involving multiphysics in multiscale. This study mainly focuses on the mechanism of mist formation and its correlation with micro-explosion. There are still many questions yet to be answered, such as the direct measurement of the vapor concentration and the quantitative analysis of the vaporization and condensation. Numerical simulation or theoretical analysis of the process will also be important to understand the mechanism. An insightful understanding of the process will be helpful for relevant applications of micro-explosion of emulsion droplets.

\section*{Declaration of Competing Interest}
None.

\section*{Acknowledgements}

This work was supported by the National Natural Science Foundation of China (Grant Nos.\ 51976133, 51676137, and 51921004).

\section*{Supplementary materials}
Supplementary material associated with this article can be found, in the online version, at doi: xxxxxx.
Movie 1: Mist formation during the intense micro-explosion of an emulsion droplet, corresponding to Figure \ref{fig:fig02}a.
Movie 2: Mist formation during the weak micro-explosion of an emulsion droplet, corresponding to Figure \ref{fig:fig02}b.
Movie 3: Mist formation during the puffing of an emulsion droplet, corresponding to Figure \ref{fig:fig02}c.
Movie 4: Mist formation during the micro-explosion of an emulsion droplet at the surfactant content of 0.5 vol\%, corresponding to Figure \ref{fig:fig14}f.

\bibliography{MicroExplosionMist}

\end{document}